\documentclass[fleqn,12pt,twoside]{article}
\usepackage[headings]{espcrc1}

\readRCS
$Id: espcrc1.tex,v 1.2 2004/02/24 11:22:11 spepping Exp $
\usepackage{graphicx}
\usepackage[figuresright]{rotating}

\newcommand{\AmS}{{\protect\the\textfont2
  A\kern-.1667em\lower.5ex\hbox{M}\kern-.125emS}}

\title{Study of the Hindrance Effect in Sub-barrier
       Fusion Reactions}

\author{Masahiro~Notani\address[ND]{Department of Physics,
        University of Notre Dame,
        Notre Dame, IN 46556, U.S.A.},
        P.~Davies\address[SR]{University of Surrey, Guildford, Surrey, GU2 7XH, U.K.},
        B.~Bucher\addressmark[ND],
        X.~Fang\addressmark[ND],
        L.~Lamm\addressmark[ND],
        C.~Ma\addressmark[ND],
        E.~Martin\addressmark[SR],
        W.~Tan\addressmark[ND],
        X.D.~Tang\addressmark[ND],
        S.~Thomas\addressmark[ND]\addressmark[SR] and
        C.L.~Jiang\address[ANL]{Physics Division, Argonne National Laboratory, Argonne, IL 60439, U.S.A.}
}
       
\runtitle{Hindrance effect in sub-barrier fusion reactions between light nuclei}
\runauthor{$M.$ $Notani$, $et$ $al$.}

\begin{document}

% typeset front matter
\maketitle

\begin{abstract}
We have measured the fusion cross sections of
the $^{12}$C($^{13}$C, $p$)$^{24}$Na reaction through
off-line measurement of the $\beta$ decay of $^{24}$Na
using the $\beta$-$\gamma$ coincidence method.
Our new measurements
in the energy range of ${E}_{c.m.}$ = 2.6$-$3.0 MeV do not
show an obvious $S$-factor maximum but a plateau.
Comparison between this work and various models
is presented.
\vspace{1pc}
\end{abstract}

%\section{Introduction}

Heavy-ion fusion reactions between light nuclei such as
carbon and oxygen isotopes have been studied because of
their significance for a wide variety of
stellar burning processes,
such as supernovas and superbursters~\cite{Gasq05}.
It has been suggested that
stars with intermediate mass or 8$-$10 of solar mass
may cause a shock wave
by carbon ignition under electron degeneracy conditions
and is observed as a supernova explosion~\cite{Garc97}. %{Arne69,Garc97}.
Superbursts on accreting neutron stars have been related
to carbon burning, for which accurate reaction rates
are required to understand
the theoretical models~\cite{Cumm06}.
The temperatures for these carbon burning processes
range from 0.8 to 1.2$\times {10}^{9}$ K,
corresponding to ${E}_{c.m.}$ = 1$-$3 MeV.

The fusion reactions have only been measured in the laboratory down
to energies that are much higher than those of astrophysical interest
so far. Optical models have been widely used to fit the experimental data,
and to estimate the reaction cross sections or the astrophysical
$S$-factor: $S(E)=E\sigma(E)\exp(2\pi\eta)$ at the lower energies
by extrapolation. Much effort has been devoted to measure
the fusion cross section such as $^{12}$C+$^{12}$C reaction at
astrophysical energies by charged particle and $\gamma$-ray
spectroscopy~\cite{Barr06}. However, the extrapolation is still
needed to estimate the reaction rate for nucleosynthesis simulation,
and the extrapolation is difficult due to the large resonant structure.

Recently, a hindrance effect in low-energy heavy-ion fusion reactions
was discovered in medium-heavy nuclei system,
such as $^{64}$Ni+$^{64}$Ni~\cite{Jiang02},
and the systematics show that the hindrance
could affect the astrophysically important fusion reactions~\cite{Jiang07}.
Therefore the measurement of fusion cross sections
in low-energy light-ion fusion reactions
is required to verify the hindrance effect.

We report an experiment to measure the cross section of $^{13}$C+$^{12}$C
reaction at extreme sub-barrier energies.
This reaction was investigated
because unlike the $^{12}$C+$^{12}$C reaction,
it does not show the large resonant structure
but relatively smooth excitation function.

%\section{Experimental Setup}

%%%%%%%%%%%%%%%%%%%%%%%%%%%%%%%%%%%%%%%%%%%%%%%%
% Experimental Setup                           %
%%%%%%%%%%%%%%%%%%%%%%%%%%%%%%%%%%%%%%%%%%%%%%%%

The $^{13}$C beam was provided by the 10-MV
FN Tandem Van de Graaff Accelerator
at the University of Notre Dame.
To obtain an intense beam, we used
a gas stripper system in the accelerator.
The charge-state distribution of $^{13}$C ions
after the gas stripper was studied and
2$^{+}$ state ions were selected.
The beam energies were determined by measuring the
magnetic field in the analyzing magnet after the
accelerator.
The $^{13}$C beam was impinged on a 1-mm thick carbon target.

%----------------------------------------------%
% FIGURE 1: Experimental Setup                 %
%----------------------------------------------%

\begin{figure}   %width=10mm,pt,height=XX,angle=45,...\rotatebox{}
\begin{minipage}[t]{77mm}
\includegraphics[scale=0.55]{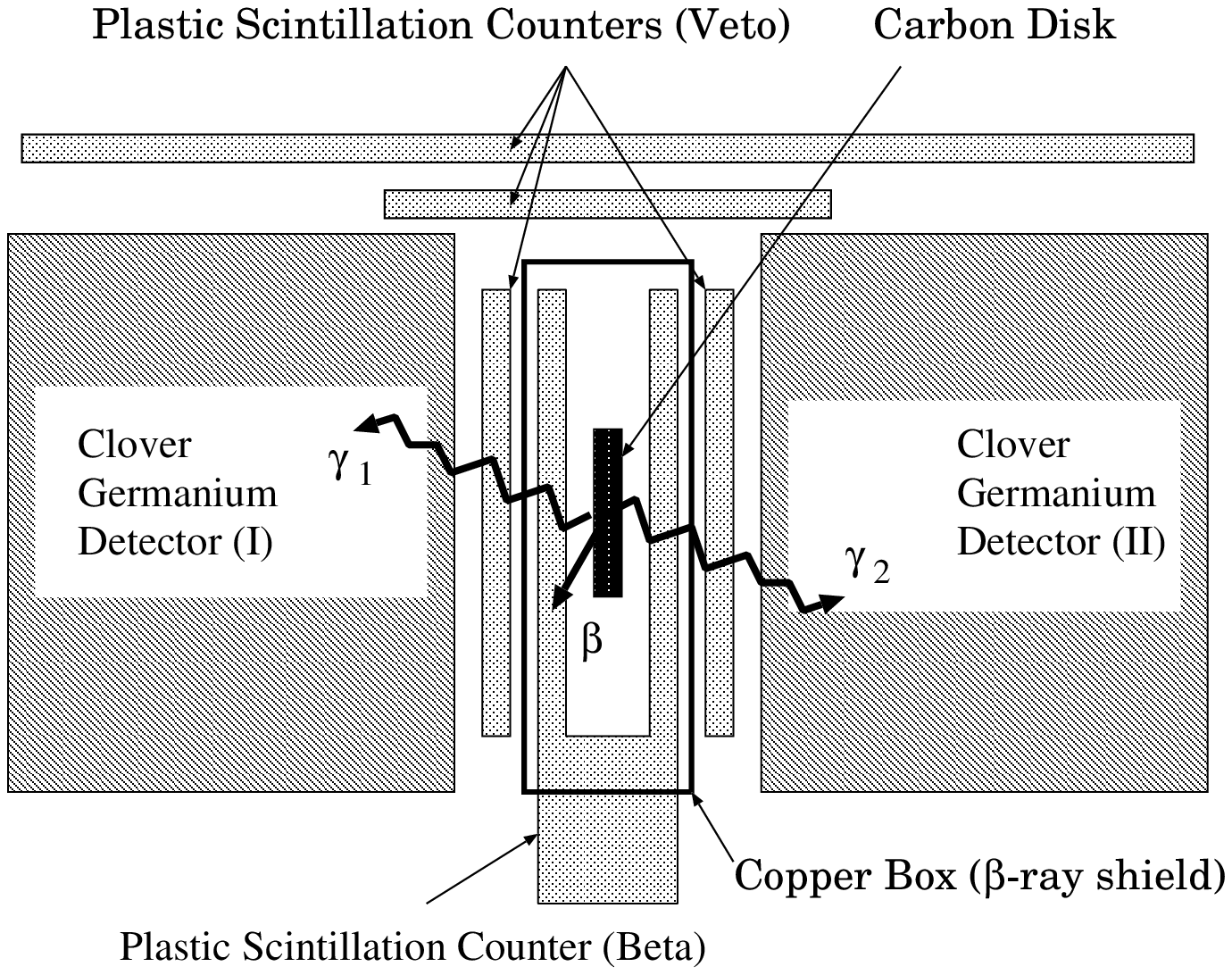}% Here is how to import EPS art
~\vspace{-10mm}\\
\caption{\label{fig:setup} Schematic view of the
experimental setup.
~\vspace{-10mm}
}
\end{minipage}
\hspace{\fill}
\begin{minipage}[t]{77mm}
\includegraphics[scale=0.42]{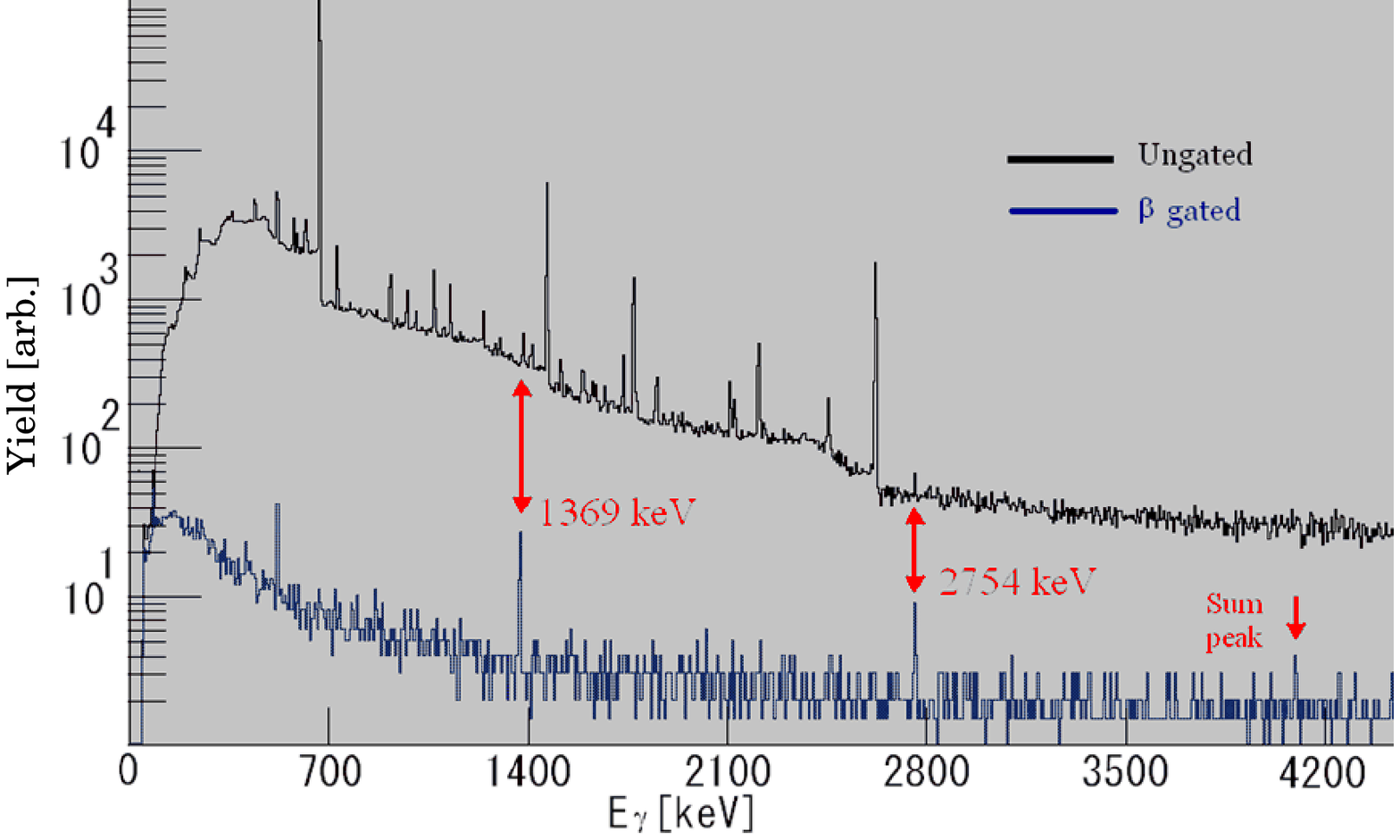}% Here is how to import EPS art
~\vspace{-10mm}\\
\caption{\label{fig:spectra}Comparison of two $\gamma$-ray spectra
obtained with and without the $\beta$-ray gate.}
\end{minipage}
\end{figure}

The cross section for the $^{12}$C($^{13}$C, $p$)$^{24}$Na reaction
has been determined through the measurement of
$\beta$-decay rate of $^{24}$Na ($T_{1/2}$ = 14.9 hrs)
using the $\beta$-$\gamma$ coincidence method.
The irradiated carbon target was transported
to a detection system for the $\beta$-decay measurement.
Figure 1 shows the experimental setup of the detector array
for the $\beta$-$\gamma$ coincidence measurement.
The carbon target was surrounded
by plastic scintillation counters
to detect the $\beta$ rays from $^{24}$Na.
Following the $\beta$ decay of $^{24}$Na,
two $\gamma$ rays (1369 keV and 2754 keV) are emitted
from the excited $^{24}$Mg.
The $\beta$-delayed $\gamma$ rays were detected
by two Ge clover detectors.
The typical spectra are shown in Fig.~\ref{fig:spectra}.
In addition, cosmic rays were vetoed by an array of
four plastic scintillation counters.
The detection system was shielded with a 7-cm thick lead.

The thick target yield was obtained by normalizing the efficiency corrected
$\beta$-gated $\gamma$-ray yield to the total incident $^{13}$C charge.
From the thick-target excitation function,
the differential yield $dY/dE$ was determined
% by graphical differentiation of the experimental data,
and the cross section was 
calculated using the following equation,
%obtained with multiplication
%of the appropriate stopping power
%as given by the $SRIM$ code and the fitting function~\cite{Barr06}.
%The cross section
%is calculated using the following equation,
%
\begin{eqnarray}
 \sigma (E) & = & \frac{{M}_{T}}{f{N}_{A}}\frac{dY}{dE}\frac{dE}{d(\rho X)},
\end{eqnarray}
where 
${M}_{T}$ is the molecular weight of the target,
$f$ is the molecular fraction of the target nucleus of interest,
${N}_{A}$ is Avogadro's number and ${dE}/{d(\rho X)}$ is
the stopping power
given by the $SRIM$ code~\cite{Barr06}.
% of the target as a function of the incident particle energy.

In order to deduce the total cross section for compound nucleus formation
or the fusion reaction cross section of $^{12}$C+$^{13}$C,
the theoretical branching ratio for
the proton emission channel
was used as correction factor,
as discussed in Ref.~\cite{Daylas76}.
%as shown in Ref.~\cite{Daylas76}.

%----------------------------------------------%
% FIGURE : Astrophysical S factor              %
%----------------------------------------------%
\begin{figure}   %width=10mm,pt,height=XX,angle=45,...\rotatebox{}
\begin{minipage}[t]{97mm}
\includegraphics[origin=c,angle=90,scale=0.45]{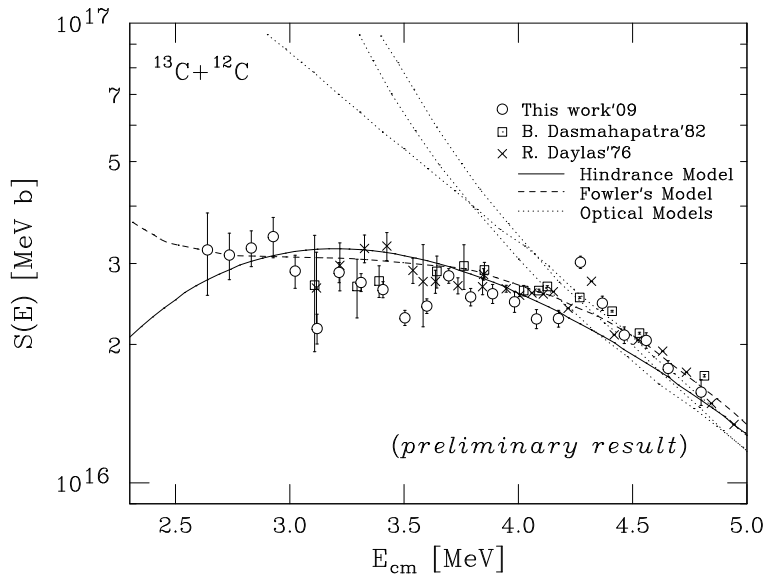}% Here is how to import EPS art
%~\vspace{-55mm}\\
\end{minipage}
~\vspace{-70mm}\\
\begin{minipage}[t]{97mm}
\end{minipage}
\hspace{\fill}
\begin{minipage}[t]{57mm}
\caption{\label{fig:sfact} Astrophysical $S$-factor $S$($E$) as a function of the center-of-mass energy $E$, derived from the fusion cross sections,
obtained from the previous experiments~\cite{Daylas76,Dasma82} and this work.}
\end{minipage}
\end{figure}

%\section{RESULT and DISCUSSION}

Figure~\ref{fig:sfact} shows the astrophysical $S$-factor
obtained from this experiment.
The result agrees with that of
the two previous measurements in the energy region from
3.2-5.0 MeV.
We have obtained 5 new data points
in the range of ${E}_{c.m.}$ = 2.6$-$3.0 MeV,
which overlaps the important energy range for the carbon burning.
%which region is astrophysically important.
Within this energy range, 
our data do not show an obvious $S$-factor maximum but a plateau.
%no obvious $S$($E$) maximum has been seen.
%
The optical models (dotted curves) with Woods-Saxon type
potential~\cite{Croz74} reproduces the experimental data
only at energies above 4 MeV.
The Fowler's model (dashed curve)
which employs a repulsive potential~\cite{Daylas76,Fowler75}
explains the plateau (${E}_{c.m.}$ $<$ 4 MeV)
and predicts an increase at very low-energy region (${E}_{c.m.}$ $<$ 2.5 MeV).
The hindrance model (solid curve)~\cite{Jiang07}
also shows a good agreement to the experimental data
and predicts a decrease at lower energies.
It should be noted that the parameters of hindrance model
were slightly modified by fitting with the present data.
%By fitting with the present data,
%the parameters for the hindrance model were modified
%and the curve is slightly flatter than the previous.
%
In order to check the difference between the extrapolations
from Fowler's model and hindrance model,
measurements at even lower energies are urgently needed.

\end{document}